\documentclass{ws-ijmpa}

\begin{document}

\markboth{Shimin Yang and Bo-Qiang Ma} {Lorentz violation in
three-family neutrino oscillation}

\catchline{}{}{}{}{}
%
%%%%%%%%%%%%%%%%%%%%%%%%%%%%%%%%%%%%%%%%%%%%%%%%%%%%%%%%%%%%%%%%%%%%

\title{LORENTZ VIOLATION IN THREE-FAMILY NEUTRINO OSCILLATION}

\author{SHIMIN YANG}

\address{School of Physics and State Key Laboratory of Nuclear Physics and Technology,\\ Peking University, Beijing, 100871, China}
%{yshmphys@pku.edu.cn}

\author{BO-QIANG MA}

\address{School of Physics and State Key Laboratory of Nuclear Physics and Technology,\\ Peking University, Beijing, 100871, China\\
mabq@phy.pku.edu.cn}

\maketitle

\begin{history}
\received{Day Month Year}
\revised{Day Month Year}
\end{history}

\begin{abstract}
We analyze the consequences of Lorentz violation (LV) to
three-generation neutrino oscillation in the massless neutrino
sector. We present a general formalism of three-family neutrino
oscillation with neutrino flavor states being mixing states of
energy eigenstates. It is also found that the mixing parts could
strongly depend on neutrino energy by special choices of Lorentz
violation parameters. By confronting with the existing experimental
data on neutrino oscillation, the upper bounds on LV parameters are
derived. Because the oscillation amplitude could vary with the
neutrino energy, neutrino experiments with energy dependence may
test and constrain the Lorentz violation scenario for neutrino
oscillation.

\end{abstract}

\ccode{PACS numbers: 11.30.Cp, 12.15.Ff, 13.15.+g, 14.60.Pq}

\section{Introduction}
\label{intro}

Lorentz invariance has been a fundamental principle in physics since
Einstein established special relativity in $1905$. In the last
century, numerous experiments provided precise verifications of
Lorentz invariance, and most current data are consistent with this
symmetry. Over the last decade there is a growing interest in
studying Lorentz invariance violation in the physics society. Though
no theoretical model predicts Lorentz violation conclusively, there
have been some theoretical suggestions that Lorentz invariance may
not be an exact symmetry at all energies. In particular, many works
to describe the force of gravity within the context of a quantum
theory imply the breaking of Lorentz symmetry (including string
theory~\cite{Kostelecky:1989prd,Ellis:1999gr}, warped brane
worlds\cite{Burgess:2002jhep}, and loop quantum
gravity\cite{Gambini:1999prd}). Other high energy models of space
structure could also contain Lorentz
violation\cite{Douglas:2001rev}. Even if Lorentz symmetry is broken
at high energy, there can still be an attractive infrared fixed
point\cite{Chadha:1983nucl}. So we can still get an approximate
Lorentz invariant world at low energy range. Besides above
theoretical motivations, many other models, such as emergent gauge
bosons\cite{Kraus:2002prd,Jenkins:2004prd}, varying
moduli\cite{Damour:1994Nucl}, ghost
condensate\cite{Arkani:2004jhep}, space-time varying
coupling\cite{Kostelecky:2003prd,bertolami:2004prd}, or varying
speed of light cosmology\cite{Moffat:1993int,Magueijo:2003rept} also
incorporate Lorentz violation. Whether Lorentz symmetry is perfectly
unbroken under all conditions is still an important theoretical
question.

Lorentz invariance violation was proposed as a solution to two
important experimental problems, i.e., the observation of TeV
photons and of cosmic ray events above the GZK
cutoff\cite{Coleman:1998ph,Coleman:1999prd,Kifune:1999astro,Protheroe:2000plb,Piran:2001ph},
for a review see Ref.~\refcite{Wolfgang:2008ar}. In the first case,
it is hard to observed ultrahigh energy photons coming from ulterior
galaxies since cosmic gamma rays with energy above $10$ TeV should
interact with cosmic infrared background photons and convert into
electron-position pairs. However, the High Energy Gamma Ray
Astronomy (HEGRA) satellite detected $24$ TeV gamma rays from
Markarian 501\cite{Aharonian:1999as}. In the second case, ultra high
energy cosmic rays interact with cosmic microwave background photons
and produce pions. The cosmic rays lose energy through this process
until they pass below the GZK cutoff energy~$10^{19}$
eV\cite{Greisen:1966prl,Zatsepin:1966jetp}. However, there has been
report that the cosmic ray spectrum extends beyond this
energy\cite{Takeda:1998prl}. Lorentz violation could give a
preparative solution to the two problems because the threshold
energy at which the cutoff occurs could be altered by modifying
special relativity, though it is not the only way to explain these
problems. We also aware\cite{Xiao:2008ar} that recent
HiRes\cite{Abbasi:2007ar} and Pierre Auger\cite{Pierre:2007sc}
measurements of the ultrahigh energy cosmic rays show a sharp
suppression around the energy of $10^{19}$ eV, which is consistent
with the expected cutoff energy. If the phenomenon of GZK cutoff is
confirmed by further experiments, Lorentz invariance will still be a
good symmetry in the area of ultra high energy cosmic ray. Therefore
one would need to look at other possibilities for the breakdown of
Lorentz invariance.

Neutrinos provide another interesting laboratory for studying the
possibility of Lorentz violation. If Lorentz invariance is violated
by quantum gravity, the natural scale one would expect a strong
Lorentz violation is the Planck energy of $10^{19}$
GeV\cite{Mattingly:2005living}. But the attainable energy of
accelerator is of TeV scale. So it is difficult to directly detect
Planck scale Lorentz violation in the laboratory. However, there
might be a small amount of Lorentz violation at lower energies if
Lorentz symmetry is violated at Planck scale. Neutrinos offer a
promising possibility to study Lorentz violation that may exist at
the low-energy as the remnants of Planck-scale
Physics\cite{Mewes:2004hep}. Neutrino oscillation is an important
problem in neutrino physics. To explain this kind of phenomena, the
conventional scenario is to assume that neutrinos have masses. In
this assumption, there is a spectrum of three or more neutrino mass
eigenstates and the flavor state is the mixing state of mass
eigenstates\cite{MNS:1962prog,Pontecorvo:1967zh}. Coleman and
Glashow pointed out that neutrino oscillation can take place even
for massless neutrinos if Lorentz invariance is violated in the
neutrino sector\cite{Coleman:1997plb}. In
Ref.~\refcite{Coleman:1999prd}, they assumed that the maximum
attainable speed of a particle depends on its identity and that the
flavor states are the mixing states of speed eigenstates, and
analyzed neutrino oscillation quantitatively. In
Ref.~\refcite{Barger:2000prl} a two-generation model to study
neutrino oscillations is established. This model involves a mass
term and a single nonzero coefficient for Lorentz violation.
Kosteleck\'{y} and Mewes presented a general formalism for
violations of Lorentz and CPT symmetry in three-family massive
neutrino oscillation\cite{Kostelecky:2004prd}. In their calculation
the mixing part has the same form comparing with that of the
conventional massive neutrino scenario because of their assumption
that neutrino are massive.  They also built a special model called
bicycle model in massless neutrino sector\cite{Kostelecky:2004prd2},
in which the Lorentz violation parameters are direction-dependent.
Ref.~\refcite{Katori:2006prd} built a three-neutrino massive model
with Lorentz-violating terms. All classes of neutrino data are
descried in this model, including LSND oscillation data. Grossman
{\it et al.} studied the interactions between the neutrino and the
Goldstone boson of spontaneous Lorentz violation, and proposed a
novel dynamic effect of Goldstone-\v{C}erekov radiation, where
neutrinos moving with respect to a preferred rest frame can
spontaneously emit Goldstone bosons\cite{Grassman:2005prd}. Arias
{\it et al.} analyzed the consequences of Lorentz violation in the
massless neutrino sector by deforming the canonical
anti-communication relations for the fields \cite{Arias:2007plb}.
Morgan {\it et al.} analyzed atmospheric neutrino oscillations at
high energy by modified dispersion relations and placed bounds on
the magnitude of this type of Lorentz invariance
violation\cite{Morgan:2007as}.

In this paper, we study Lorentz violation contribution to neutrino
oscillation. In our calculation we assume that neutrinos are
massless and that the neutrino flavor states are mixing states of
energy eigenstates. We calculate neutrino oscillation probabilities
by the effective theory for Lorentz violation, which is usually
called the standard model extension
(SME)\cite{Kostelecky:1997prd,Kostelecky:1998prd}. In our work, the
mixing angles for neutrinos are functions of Lorentz violation
parameters.

The structure of this paper is as follows. In
Sect.~\ref{sec:theory}, we figure out the neutrino oscillation
probabilities by the effective theory for Lorentz violation. In
Sect.~\ref{sec:model}, we introduce some specific models and give
the numerical values for LV parameters by comparing our theoretical
oscillation probabilities with experimental results. Then we analyze
the dependence of the new oscillation equation on neutrino energy
and neutrino propagation length. Remarks and conclusions are given
in Sect.~\ref{sec:conclu}.

\section{Theoretical framework for Lorentz violation and neutrino oscillation
probabilities}
\label{sec:theory}

In the conventional massive neutrinos scenario, neutrino mass
eigenstates are components of neutrino flavor states. Neutrinos
change from one flavor to another during the propagation, because
the phase of time for each mass eigenstate is different. However,
Lorentz invariance violation may be another origin for neutrino
oscillation. In this work, we analyze the consequences of Lorentz
violation in the massless neutrino sector. Different from the
conventional massive neutrino model, neutrino flavor states are
mixing states of eigenenergy in our calculation. The bounds on LV
parameters will be given by comparing our calculation with the
experimental results. In this part we will figure out the neutrino
oscillation probability. The general and detailed calculation method
for neutrino oscillation was proposed in
Ref.~\refcite{Kostelecky:2004prd}. However, for the integrity of the
article, we display the entire calculation for neutrino oscillation
probabilities.

In the framework of standard model extension, we consider the
Lagrangian \cite{Kostelecky:1998prd,Kostelecky:2001prd} for neutrino
sector given by
\begin{equation}
\label{Lv:equ:Lagrange}
\mathcal{L}=\frac{1}{2}i\overline{\nu}_{_A}\gamma^{\mu}\overleftrightarrow{\partial_{\mu}}\nu_{_B}\delta_{_{AB}}
+\frac{1}{2}ic_{_{AB}}^{\mu\nu}\overline{\nu}_{_A}\gamma^{\mu}\overleftrightarrow{\partial^{\nu}}\nu_{_B}
-a_{_{AB}}^{\mu}\overline{\nu}_{_A}\gamma^{\mu}\nu_{_B},~~
\end{equation}
where $\mu$, $\nu$ are spinor indices and $A$, $B$ are flavor
indices. $c_{_{AB}}^{\mu\nu}$ and $a_{_{AB}}^{\mu}$ are LV
parameters. In Eq. (\ref{Lv:equ:Lagrange}), the first term is
consistent with the minimal standard model; the second and third
terms describe the contribution from Lorentz violation, which denote
CPT even term and CPT odd term respectively. For the convenience of
calculation, we transform the lagrangian as
\begin{equation}
\label{Lv:equ:trLagrange}
\mathcal{L}'=i\hspace{1pt}\overline{\nu}_{_A}\gamma^{\mu}\partial_{\mu}\nu_{_B}\delta_{_{AB}}
+ic_{_{AB}}^{\mu\nu}\overline{\nu}_{_A}\gamma^{\mu}\partial^{\nu}\nu_{_B}
-a_{_{AB}}^{\mu}\overline{\nu}_{_A}\gamma^{\mu}\nu_{_B},~~
\end{equation}
where $\mathcal {L}'$ equals to $\mathcal {L}$ in quantum field
theory, because divergence terms have no contribution to the action.
The Euler-Lagrange equation of motion for neutrinos can be written
as
\begin{equation}
\label{Lv:equ:motion}
 i\gamma^{0}\partial_{0}\nu_{_A}+i\gamma^{i}\partial_{i}\nu_{_A}+ic_{_{AB}}^{\mu\nu}\gamma^{\mu}\partial^{\nu}\nu_{_B}-a_{_{AB}}^{\mu}\gamma^{\mu}\nu_{_B}=0.
\end{equation}
A new motion equation can be obtained by multiplying
Eq.~(\ref{Lv:equ:motion}) with the matrix $\gamma^{0}$. Comparing
the new equation with the conventional equations of motion
$(i\delta_{_AB}\partial_{0}-\mathcal{H}_{_AB})\nu_{_A}=0$, we obtain
the Hamiltonian for neutrinos
\begin{eqnarray}
\label{Lv:equ:Hamilton}
\mathcal{H}=-i\gamma^{0}\gamma^{i}\partial_{i}-ic_{_{AB}}^{\mu\nu}\gamma^{0}\gamma^{\mu}\partial^{\nu}+a_{_{AB}}^{\mu}\gamma^{0}\gamma^{\mu}.
\end{eqnarray}
Note the unconventional time-derivative term in
Eq.~(\ref{Lv:equ:motion}) has been included in
Eq.~(\ref{Lv:equ:Hamilton}). The above discussion is within the
context of quantum field theory. Now we transform the description
into quantum mechanics and treat the differential operator as
momentum operator to study neutrinos with a fixed momentum. In this
paper, we focus on the upper bounds of Lorentz violation
contribution to neutrino oscillation and so far right-handed
neutrinos or left-handed antineutrinos have not been detected
experimentally. So we assume that neutrinos are massless. Our
calculation is based on three-generation model. With the basic
vector $(u_{_L}(p) ,\hspace{0.03in} v_{_R}(-p))^{T}$, the
Hamiltonian matrix for neutrinos can be given by
\begin{equation}
\label{Lv:equ:mHamilton}
\newcommand{\DF}[2]{{\displaystyle\frac{#1}{#2}}}
 H_{_{AB}}= \begin{pmatrix}
|\overrightarrow{p}|\delta_{_{AB}}+c_{_{AB}}^{\mu
\nu}\DF{p_{\mu}p_{\nu}}{|\overrightarrow{p}|}+a_{_{AB}}^{\mu}\DF{p_{\mu}}{|\overrightarrow{p}|}
& 0\\0 & |\overrightarrow{p}|\delta_{_{AB}}+c_{_{AB}}^{\mu
\nu}\DF{p_{\mu}p_{\nu}}{|\overrightarrow{p}|}-a_{_{AB}}^{\mu}\DF{p_{\mu}}{|\overrightarrow{p}|}
\end{pmatrix}.
\end{equation}

Substituting Eq.~(\ref{Lv:equ:mHamilton}) into
Eq.~(\ref{Lv:equ:motion}) the general dynamical equation for
neutrinos could be obtained. Then we can study the dynamical
character of neutrinos by calculating the dynamical equation.
However, we are more interested in neutrino oscillation probability
in this work. So we just need to figure out the eigenenergy for
neutrinos by diagonalizing the Hamiltonian. In Eq.
(\ref{Lv:equ:mHamilton}), the up-diagonal term is the Hamiltonian
for left-handed neutrinos and the down-diagonal term is the
Hamiltonian for right-handed antineutrinos. We just study the
left-handed neutrino part because the two terms have the similar
form. We simplify the Hamiltonian $H_{_{AB}}$ as
\begin{equation}
\label{Lv:equ:effHamilton}
\newcommand{\DF}[2]{{\displaystyle\frac{#1}{#2}}}
h_{_{AB}}=|\overrightarrow{p}|\delta_{_{AB}}+c_{_{AB}}^{\mu
\nu}\DF{p_{\mu}p_{\nu}}{|\overrightarrow{p}|}+a_{_{AB}}^{\mu}\DF{p_{\mu}}{|\overrightarrow{p}|},
\end{equation}
where $A$, $B$ $=$ $e$, $\mu$, $\tau$. To figure out neutrino
eigenenergy we diagonalize the Hamiltonian matrix
(\ref{Lv:equ:effHamilton}) by a $3\times 3$ matrix $U$. Note that
$U$ is a unitary matrix. The eigenenergy matrix is
\begin{equation}
\label{Lv:equ:eigenenergy}
E_{_{IJ}}=U_{_{IA}}^{^\dagger}h_{_{AB}}U_{_{BJ}},
\end{equation}
where $E$ is a diagonalized matrix. The eigenenergy can be labeled
as $E_{_I}$, where $I$ $=$ $1$, $2$, $3$. Neutrino energy
eigenstates are the linear combination of flavor eigenstates because
matrix $U$ is unitary:
\begin{equation}
\label{Lv:equ:mixingstate}
|\nu_{_I}\rangle=(U^{^\dagger})_{_{IA}}|\nu_{_A}\rangle,
\end{equation}
where $I$ and $A$ represent different eigenstates and flavor
eigenstates respectively. Because the matrix $U$ is unitary,
Eq.~(\ref{Lv:equ:mixingstate}) can be transformed as
\begin{equation}
\label{Lv:equ:mixingstate2} |\nu_{_A}(t)\rangle=\sum_{_{IB}}
(U^{^\dagger})_{_{IA}}^{\ast}e^{{-iE_{_I}t}}(U^{^\dagger})_{_{IB}}|\nu_{_B}\rangle.
\end{equation}
Neutrinos propagate at the speed of light approximately because we
have assumed that neutrinos are massless and that LV parameters is
small. When the propagation length is $L$, the flavor state can be
written as
\begin{equation}
\label{Lv:equ:mixingstate3} |\nu_{_A}(L)\rangle=\sum_{_{IB}}
(U^{^\dagger})_{_{IA}}^{\ast}e^{-iE_{_I}L}(U^{^\dagger})_{_{IB}}|\nu_{_B}\rangle.
\end{equation}

>From Eq.~(\ref{Lv:equ:mixingstate3}) and the unitary of $U$, the
oscillation probability can be expressed as
\begin{multline}
\label{Lv:equ:probability}
P(\nu_{_A}\to \nu_{_B})=\delta_{_{AB}}\\
-4\sum_{_{I>J}}\Re[(U^{^\dagger})_{_{IA}}^{\ast}(U^{^\dagger})_{_{IB}}(U^{^\dagger})_{_{JA}}^{\ast}(U^{^\dagger})_{_{JB}}]\sin^{2}(\frac{\Delta
E_{_{IJ}}}{2}L)
\\
+2\sum_{_{I>J}}\Im[(U^{^\dagger})_{_{IA}}^{\ast}(U^{^\dagger})_{_{IB}}(U^{^\dagger})_{_{JA}}^{\ast}(U^{^\dagger})_{_{JB}}]\sin^{2}(\Delta
E_{_{IJ}}L),
\end{multline}
where $\Re$ and $\Im$ denote the real and imaginary parts
respectively.

\section{Specific models and bounds on LV parameters}
\label{sec:model}

\label{sec:model} In Sect.~\ref{sec:theory}, we obtained the
analytic equations for neutrino oscillation. In this section, we
discuss three specific models. Comparing with the experimental
results, we will figure out the upper bounds on LV parameters in
different models. The Lorentz noninvariant direction-dependent
oscillations for massless neutrinos are not supported by recent
research work\cite{arXiv:0706.1085,MINOS:arxiv}. So in this part we
try to build a massless neutrino model without direction-dependent
oscillations. All the three special models in our work are
direction-independent.

In Eq.~(\ref{Lv:equ:effHamilton}), $h_{_{AB}}$ is a $3\times3$
matrix in neutrino generation space. To reduce the LV parameters, we
assume that the neutrino Hamiltonian can be simplified as
\begin{equation}
\label{Model:equ:H} h_{_{AB}}=\begin{pmatrix} E & \varepsilon & 0 \\
\varepsilon & E+\eta & \zeta \\ 0 & \zeta & E \end{pmatrix},
\end{equation}
where $E = |\overrightarrow{p}|$, $\varepsilon =
c_{e\mu}^{00}p_{_0}+a_{e\mu}^{0} = c_{e\mu}^{00}E+a_{e\mu}^{0}$,
$\zeta = c_{\mu\tau}^{00}E+a_{\mu\tau}^{0}$, and $\eta =
c_{\mu\mu}^{00}E+a_{\mu\mu}^{0}$. Lorentz violation parameters are
defined in a sun-centered inertial frame. There are 6 non-zero LV
parameters in the above specific model, and the other LV parameters
are zero.

By substituting Eq. (\ref{Model:equ:H}) into Eq.
(\ref{Lv:equ:eigenenergy}), the eigenenergy matrix $E$ and mixing
matrix $U$ could be figured out as

\begin{equation}
\label{Model:equ:E}
\newcommand{\DF}[2]{{\displaystyle\frac{#1}{#2}}}
E_{_{IJ}}=\begin{pmatrix} E \quad & 0 & 0 \\[0.1in] 0 \quad &
E+\DF{\eta}{2}-\DF{\sqrt{4\varepsilon^{2}+4\zeta^{2}+\eta^{2}}}{2} & 0 \\[0.1in] 0 \quad & 0 &
E+\DF{\eta}{2}+\DF{\sqrt{4\varepsilon^{2}+4\zeta^{2}+\eta^{2}}}{2}
\end{pmatrix},
\end{equation}

and
\begin{equation}
\label{Model:equ:U}
\newcommand{\DF}[2]{{\displaystyle\frac{#1}{#2}}}
 U^{^\dag}=\begin{pmatrix}
-\DF{\zeta}{\sqrt{\varepsilon^{2}+\zeta^{2}}} & 0 &
\DF{\varepsilon}{\sqrt{\varepsilon^{2}+\zeta^{2}}}
\\[0.3in]
\DF{\varepsilon}{\sqrt{N}} &
\DF{\eta-\sqrt{4(\varepsilon^{2}+\zeta^{2})+\eta^{2}}}{2\sqrt{N}} &
\DF{\zeta}{\sqrt{N}}
\\[0.3in]
\DF{\varepsilon}{\sqrt{M}} &
\DF{\eta+\sqrt{4(\varepsilon^{2}+\zeta^{2})+\eta^{2}}}{2\sqrt{M}} &
\DF{\zeta}{\sqrt{M}}
\end{pmatrix},
\end{equation}
where
\begin{eqnarray}
M&=&2(\varepsilon^{2}+\zeta^{2})+\frac{\eta^{2}}{2}+\frac{\eta\sqrt{4(\varepsilon^{2}+\zeta^{2})+\eta^{2}}}{2},\nonumber \\
N&=&2(\varepsilon^{2}+\zeta^{2})+\frac{\eta^{2}}{2}-\frac{\eta\sqrt{4(\varepsilon^{2}+\zeta^{2})+\eta^{2}}}{2}.
\end{eqnarray}
With the eigenenergy matrix $E$ and the unitary matrix $U$, the
oscillation probabilities can be expressed as

\begin{eqnarray}
\label{Model:equ:P}
P(\nu_{e}\to\nu_{e})&=&1 \nonumber \\
&&-\frac{4\varepsilon^{2}\zeta^{2}}{(\varepsilon^{2}+\zeta^{2})[\sqrt{2(\varepsilon^{2}+\zeta^{2})+\frac{\eta^{2}}{2}-\frac{\eta\sqrt{4(\varepsilon^{2}+\zeta^{2})+\eta^{2}}}{2}}]}\sin^{2}[(\frac{\eta}{4}-\frac{\sqrt{4\varepsilon^{2}+4\zeta^{2}+\eta^{2}}}{4})L]\nonumber \\
 & &-\frac{4\varepsilon^{2}\zeta^{2}}{(\varepsilon^{2}+\zeta^{2})[\sqrt{2(\varepsilon^{2}+\zeta^{2})+\frac{\eta^{2}}{2}+\frac{\eta\sqrt{4(\varepsilon^{2}+\zeta^{2})+\eta^{2}}}{2}}]}\sin^{2}[(\frac{\eta}{4}+\frac{\sqrt{4\varepsilon^{2}+4\zeta^{2}+\eta^{2}}}{4})L]\nonumber\\
 & &-\frac{4\varepsilon^{4}}{(\varepsilon^{2}+\zeta^{2}+\eta^{2})(4(\varepsilon^{2}+\zeta^{2})+\eta^{2})}\sin^{2}[(\frac{\sqrt{4(\varepsilon^{2}+\zeta^{2})+\eta^{2}}}{2})L],
 \nonumber \\[0.05in]
 P(\nu_{e}\to\nu_{\mu})&=&\frac{4\varepsilon^{2}}{4\varepsilon^{2}+4\zeta^{2}+\eta^{2}}\sin^{2}[(\frac{\sqrt{4\varepsilon^{2}+4\zeta^{2}+\eta^{2}}}{2})L],
 \nonumber \\[0.05in]
 P(\nu_{\mu}\to\nu_{\mu})&=&1-\frac{4\varepsilon^{2}+4\zeta^{2}}{4\varepsilon^{2}+4\zeta^{2}+\eta^{2}}\sin^{2}[(\frac{\sqrt{4\varepsilon^{2}+4\zeta^{2}+\eta^{2}}}{2})L],
 \nonumber \\[0.05in]
 P(\nu_{\mu}\to\nu_{\tau})&=&\frac{4\zeta^{2}}{4\varepsilon^{2}+4\zeta^{2}+\eta^{2}}\sin^{2}[(\frac{\sqrt{4\varepsilon^{2}+4\zeta^{2}+\eta^{2}}}{2})L],
 \nonumber \\[0.05in]
P(\nu_{e}\to\nu_{\tau})&=&\frac{4\varepsilon^{2}\zeta^{2}}{(\varepsilon^{2}+\zeta^{2})[\sqrt{2(\varepsilon^{2}+\zeta^{2})+\frac{\eta^{2}}{2}-\frac{\eta\sqrt{4(\varepsilon^{2}+\zeta^{2})+\eta^{2}}}{2}}]}\sin^{2}[(\frac{\eta}{4}-\frac{\sqrt{4\varepsilon^{2}+4\zeta^{2}+\eta^{2}}}{4})L]\nonumber
 \\& &+\frac{4\varepsilon^{2}\zeta^{2}}{(\varepsilon^{2}+\zeta^{2})[\sqrt{2(\varepsilon^{2}+\zeta^{2})+\frac{\eta^{2}}{2}+\frac{\eta\sqrt{4(\varepsilon^{2}+\zeta^{2})+\eta^{2}}}{2}}]}\sin^{2}[(\frac{\eta}{4}+\frac{\sqrt{4\varepsilon^{2}+4\zeta^{2}+\eta^{2}}}{4})L]\nonumber
 \\&
 &-\frac{4\varepsilon^{2}\zeta^{2}}{(\varepsilon^{2}+\zeta^{2})(4\varepsilon^{2}+4\zeta^{2}+\eta^{2})}\sin^{2}[(\frac{\sqrt{4\varepsilon^{2}+4\zeta^{2}+\eta^{2}}}{2})L],
 \nonumber\\
P(\nu_{\tau}\to\nu_{\tau})&=&1\nonumber \\
 & &-\frac{4\varepsilon^{2}\zeta^{2}}{(\varepsilon^{2}+\zeta^{2})[\sqrt{2(\varepsilon^{2}+\zeta^{2})+\frac{\eta^{2}}{2}-\frac{\eta\sqrt{4(\varepsilon^{2}+\zeta^{2})+\eta^{2}}}{2}}]}\sin^{2}[(\frac{\eta}{4}-\frac{\sqrt{4\varepsilon^{2}+4\zeta^{2}+\eta^{2}}}{4})L]\nonumber
 \\& &-\frac{4\varepsilon^{2}\zeta^{2}}{(\varepsilon^{2}+\zeta^{2})[\sqrt{2(\varepsilon^{2}+\zeta^{2})+\frac{\eta^{2}}{2}+\frac{\eta\sqrt{4(\varepsilon^{2}+\zeta^{2})+\eta^{2}}}{2}}]}\sin^{2}[(\frac{\eta}{4}+\frac{\sqrt{4\varepsilon^{2}+4\zeta^{2}+\eta^{2}}}{4})L]\nonumber
 \\&
 &-\frac{4\zeta^{4}}{(\varepsilon^{2}+\zeta^{2})(4\varepsilon^{2}+4\zeta^{2}+\eta^{2})}\sin^{2}[(\frac{\sqrt{4\varepsilon^{2}+4\zeta^{2}+\eta^{2}}}{2})L].
\end{eqnarray}

In Eq.~(\ref{Model:equ:P}), oscillation probabilities satisfy the
probability unity equation $\sum_{_B} P(\nu_{_A}\to\nu_{_B}) = 1$,
which is guaranteed by the unitarity of $U$. From above calculation,
we can see that not only the eigenenergy splitting $\Delta\,E_{ij}$
but also the mixing parts are functions of LV parameters. To clearly
understand the novel effect of Eq.~(\ref{Model:equ:P}), we can
analyze $P(\nu_{\mu}\to\nu_{\tau})$ with a special choice of LV
parameters. If $\eta$ is the only nonzero CPT even LV parameter,
then $\eta=c_{\mu\mu}^{00}E$, $\varepsilon=a_{e\mu}^{0}$,
$\zeta=a_{\mu\tau}^{0}$. Now both the mixing part and the
eigenenergy splitting $\Delta~E_{ij}$ are functions of neutrino
energy. In the conventional massive neutrino model, the
corresponding probability can be expressed as
$P(\nu_{\mu}\to\nu_{\tau})=\sin^{2}2\theta\cdot\,\sin^{2}[1.27\,\Delta\,m^{2}(L/E)]$,
while the mixing part $\sin^{2}2\theta$ is independent of energy. So
the amplitude of conventional massive neutrino oscillation is fixed,
and the phenomenon of oscillation disappears when the neutrino
energy is high enough because the oscillation $P$ directs to $L/E$
at high energy. But in Eq.~(\ref{Model:equ:P}), we can clearly see
that the mixing part of $P(\nu_{\mu}\to\nu_{\tau})$ is a function of
neutrino energy. So oscillation amplitude varies with neutrino
energy. By further calculation, as shown in \ref{sec:model1}, the
oscillation probability will be suppressed at high energy, which is
consistent with the conventional massive model, but the suppressed
threshold energy is much higher comparing with the conventional
massive model.

There would be different novel behaviors of the neutrino oscillation
in the presence of Lorentz violation by choosing different LV
parameters. In the following we will study three special cases with
different kinds of choosing LV parameters. By comparing with the
exiting experimental data, we can constrain LV parameters.

\subsection{Model 1 Non-diagonal terms are CPT-Odd parameters with CPT-Even parameters only in the diagonal term}
\label{sec:model1}

If we assume that $a_{e\mu}^{0}$, $a_{\mu\tau}^{0}$, and
$c_{\mu\mu}^{00}E$  are nonzero, then
\begin{equation}
\varepsilon=a_{e\mu}^{0}, \qquad \zeta=a_{\mu\tau}^{0}, \qquad
\eta=c_{\mu\mu}^{00}E.
\end{equation}
Substitute these LV parameters into Eq.~(\ref{Lv:equ:probability}).
To calculate the numerical results of LV parameter, we compare the
theoretical equations with experimental results.

KamLAND detected $P(\nu_{e}\to\nu_{e})\simeq61\%$ with neutrino
energy $E\simeq4.3$ MeV and neutrino propagation length $L\simeq180$
km \cite{Kamland:2003prl}. MINOS observed muon neutrino
disappearance with neutrino energy $E\simeq4.9$ GeV and propagation
length $L\simeq735$ km, $P(\nu_{\mu}\to\nu_{\mu})\simeq76\%$
\cite{MINOS:2006prl}. We also get the flavor change channel of muon
neutrino to tau neutrino from K2K,
$P(\nu_{\mu}\to\nu_{\tau})\simeq36\%$ with $E\simeq1.8$ GeV,
$L\simeq250$ km \cite{K2K:2005prl}.

Substituting these experimental data into
Eq.~(\ref{Lv:equ:probability}), we get three equations with three LV
parameters. Because this is a nonlinear system of equations, the
solution is not unique. But we can give the order of LV parameters
in principle: $a_{e\mu}^{0}\sim a_{\mu\tau}^{0}\sim10^{-11}$ eV and
$c_{\mu\mu}^{00}\sim10^{-20}$.

For further restriction of LV parameters, we use the experimental
results for the oscillation channel of electron neutrino to muon
neutrino. LSND has used muon sources from the decay
$\pi^{+}\to\mu^{+}+\nu_{\mu}$ to detect neutrino oscillation in the
subsequent decay of the muon through $\mu^{+}\to
e^{+}+\nu_{e}+\bar{\nu}_{\mu}$ \cite{LSND:2001prd}. This experiment
finds the oscillation $\overline{\nu}_{\mu}\to\overline{\nu}_{e}$
(with $20$ MeV $\leq E_{\nu_{\mu}}\leq 58.2$ MeV) with a probability
of $0.26\%$. However, the MiniBoone experiment, which was expected
to verify the results of LSND, has reported their first result which
does not favor the simple explanation of the LSND results based on
the two flavor neutrino oscillation \cite{MiniBooNE:2007prl}. In
this paper, we use the experimental results from K2K about the
oscillation channel $\nu_{\mu}\to\nu_{e}$ with the mixing angle
$\sin ^{2}2\theta_{\mu e}<0.13$ \cite{K2K:2006prl}. Now we plot the
oscillation probability as a function of the neutrino energy $E$,
and as a function of the path length $L$, respectively, with the LV
parameters $a_{e\mu}^{0}=1.09\times10^{-11}$ eV,
$a_{\mu\tau}^{0}=2.97\times10^{-11}$ eV, and
$c_{\mu\mu}^{00}=1.42\times10^{-20}$.
\begin{figure}[h]
\centering \scalebox{0.7}{\includegraphics[0,0][250,197]{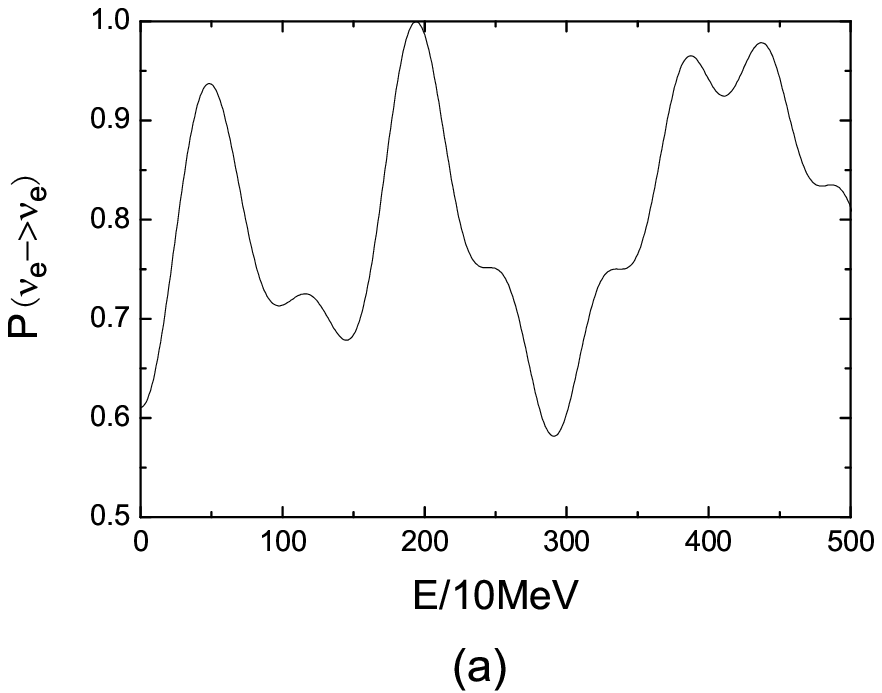}}
\scalebox{0.7}{\includegraphics[0,0][250,197]{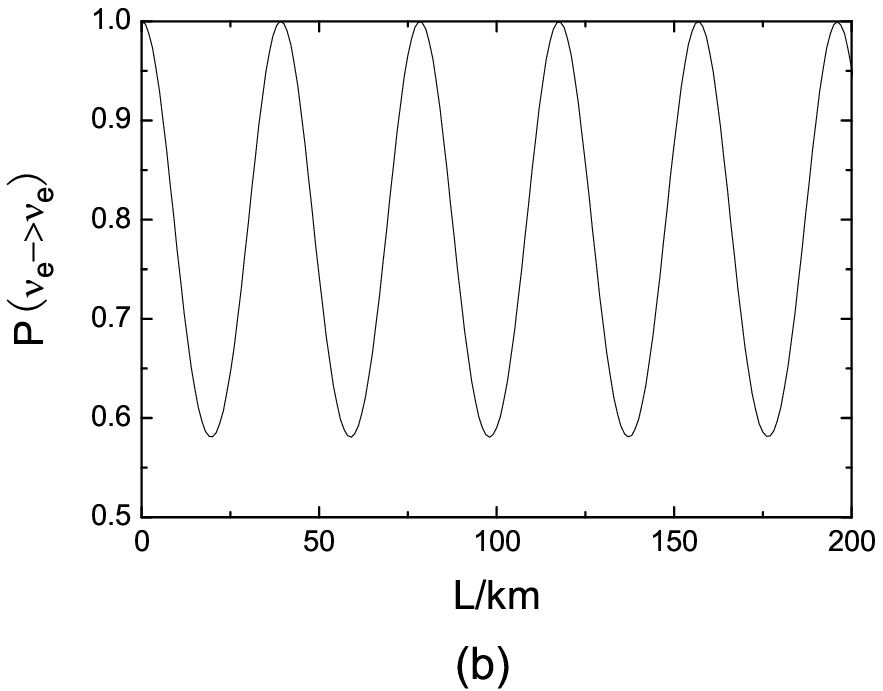}}
\caption{Neutrino oscillation probabilities for $\nu_{e}\to\nu_{e}$.
(a) The left plot shows the oscillation probability as a function of
neutrino energy with a fixed path length $L=180$ km. (b) The right
plot shows the oscillation probability as a function of path length
with a fixed energy $E=10$ MeV.} \label{fig:1}
\end{figure}
\begin{figure}[h]
\centering \scalebox{0.7}{\includegraphics[0,0][250,197]{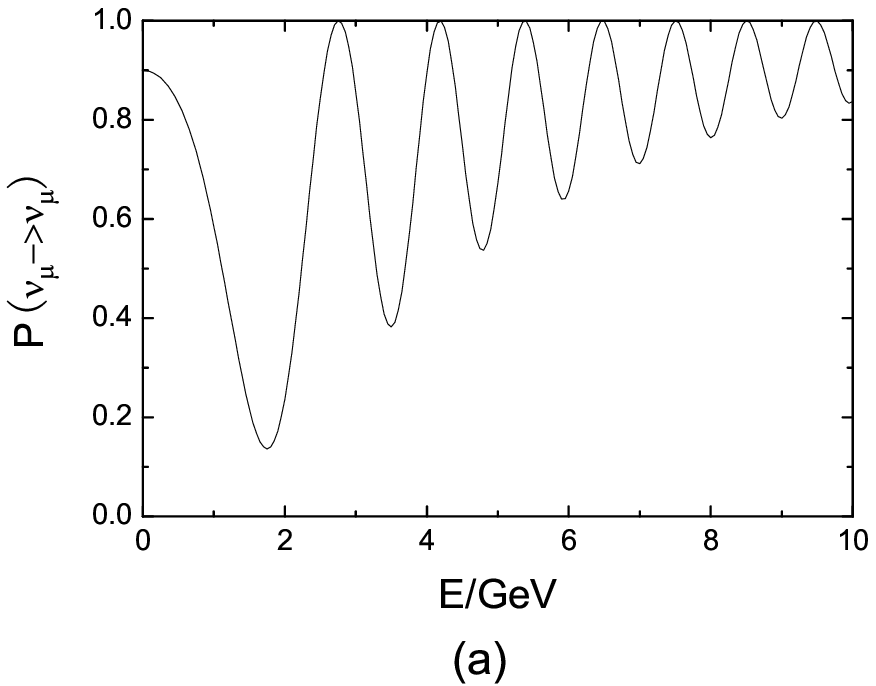}}
\scalebox{0.7}{\includegraphics[0,0][250,197]{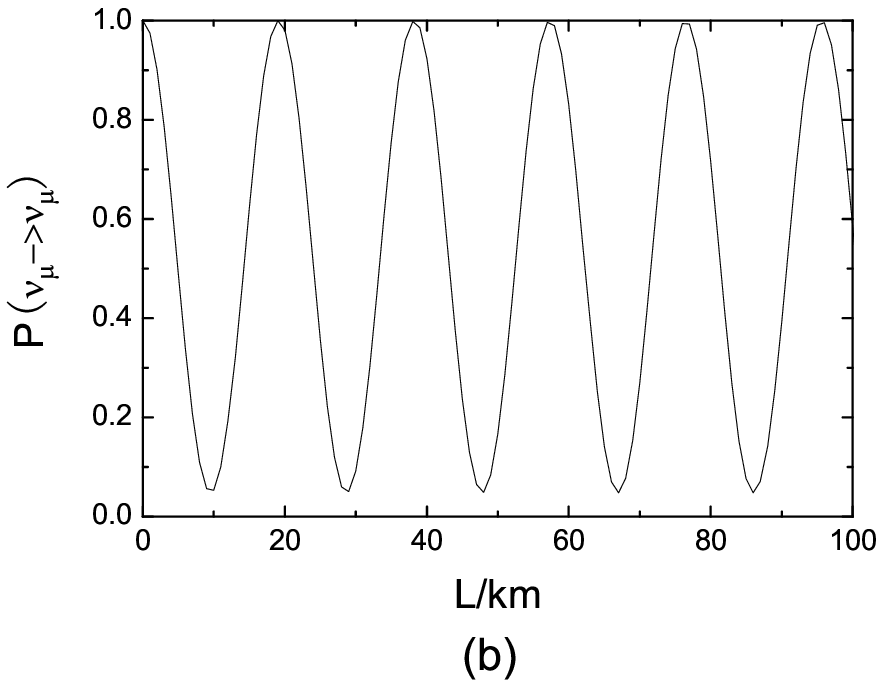}}
\caption{Neutrino oscillation probabilities for
$\nu_{\mu}\to\nu_{\mu}$. (a) The left plot shows the oscillation
probability as a function of neutrino energy with a fixed path
length $L=100$ km. (b) The right plot shows the oscillation
probability as a function of path length with a fixed energy $E=1$
GeV.} \label{fig:2}
\end{figure}

Fig.~\ref{fig:1} shows the oscillation probability for
$\nu_{e}\to\nu_{e}$ as a function of the neutrino energy $E$ with a
fixed path length $L=180$ km, and as a function of the propagation
length $L$ with neutrino energy $E=10$ MeV, respectively. Comparing
with the results of massive neutrino scenario, we find that in
Lorentz violation model electron neutrino does not have drastic
oscillation at low energy. This is different from the conventional
massive neutrino scenario. In conventional neutrino model, the
phenomena of neutrino oscillation ($\nu_{e}\to\nu_{e}$) disappears
at high energy ($E>100$ MeV with a fixed path length $L=180$ km).
However, in Fig.~\ref{fig:1} we find the existence of oscillations
at high energy. This distinction is resulted from different
dependence on neutrino energy for oscillation probability. In
addition, the mixing part is fixed in convention massive model. But
in Fig.~\ref{fig:2} we could find that the neutrino oscillation
amplitude varies with neutrino energy. These novel effects can be
tested at high energy sector. Because the dependence on path length
has the same form in two different models (massive neutrino model
and Lorentz violation model), the plots of path length dependence
have no remarkable distinction with the conventional massive
neutrino model.

\subsection{Model 2 non-diagonal terms are CPT-Even parameters with CPT-Odd parameters only in the diagonal term}
\label{sec:model2}

If we assume that $a_{\mu\mu}^{0}$, $c_{e\mu}^{00}$, and
$c_{\mu\tau}^{00}$ are nonzero, then
\begin{equation}
\varepsilon=c_{e\mu}^{00}E, \qquad \zeta=c_{\mu\tau}^{00}E, \qquad
\eta=a_{\mu\mu}^{0}.
\end{equation}

\begin{figure}
\centering \scalebox{0.7}{\includegraphics[0,0][250,197]{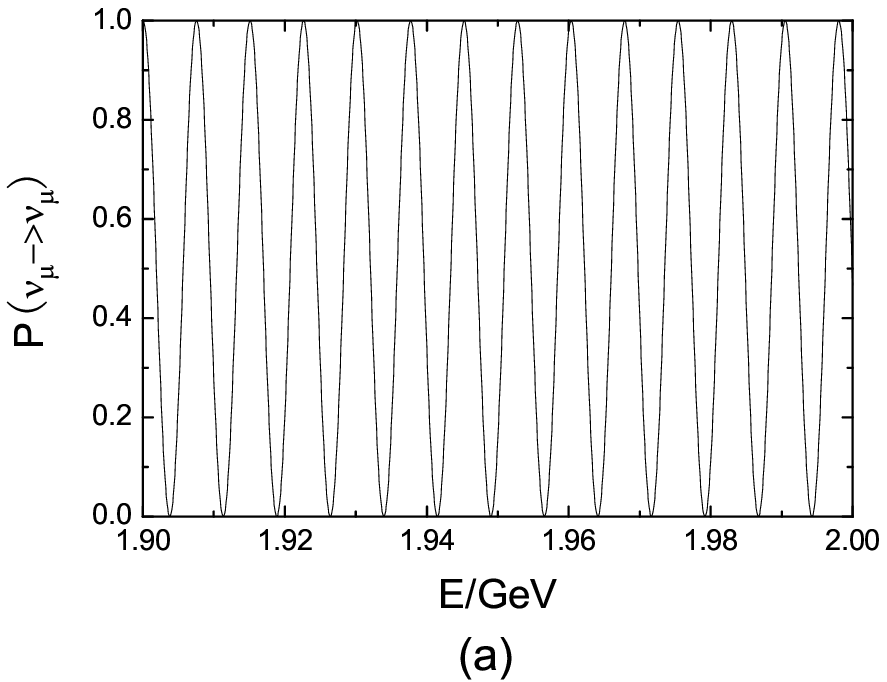}}
\scalebox{0.7}{\includegraphics[0,0][250,197]{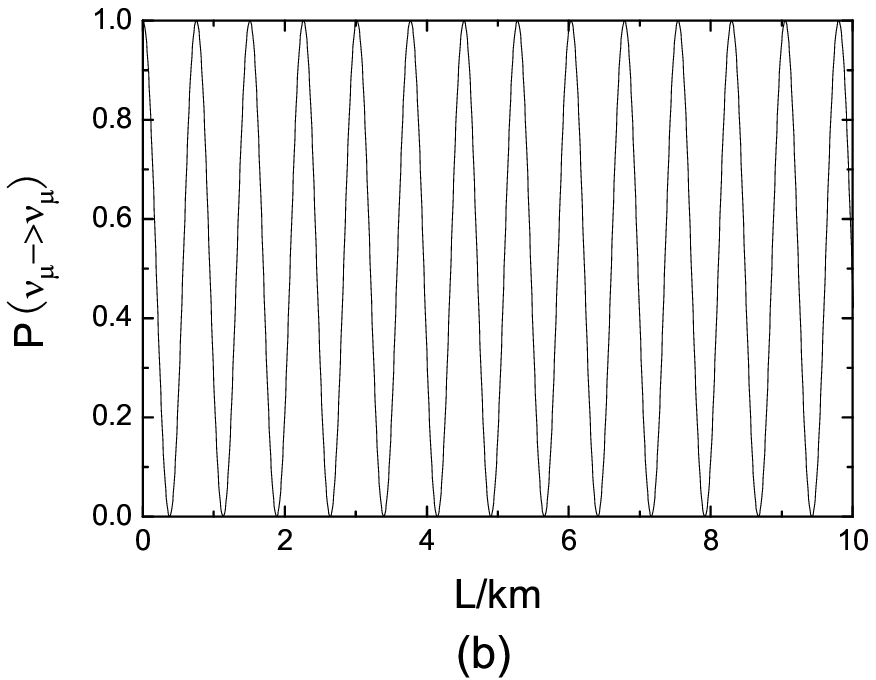}}
\caption{Neutrino oscillation probabilities for
$\nu_{\mu}\to\nu_{\mu}$. (a) The left plot shows the oscillation
probability as a function of neutrino energy with a fixed path
length $L=100$ km. (b) The right plot shows the oscillation
probability as a function of path length with a fixed energy $E=1$
GeV.} \label{fig:3}
\end{figure}

Substitute these LV parameters into Eq.~(\ref{Lv:equ:probability}).
Similarly to Model 1, we use KamLAND, K2K and MINOS experimental
results to compute the parameters. The numerical values are also
consistent with experimental results about the oscillation channel
$\nu_{\mu}\to\nu_{e}$ from K2K. Fig.~\ref{fig:3} shows the
oscillation probability of the channel $\nu_{\mu}\to\nu_{\mu}$ as a
function of the neutrino energy $E$ and as a function of the
propagation length $L$ respectively with LV parameters
$a_{\mu\mu}^{0}=1.58\times10^{-12}$ eV,
$c_{e\mu}^{00}=2.93\times10^{-19}$, and
$c_{\mu\tau}^{00}=7.68\times10^{-19}$.

The order of LV parameters $c_{_{AB}}$ is larger than that of Model
1, so the oscillation periods of the changing probabilities against
neutrino energy and path length are much smaller than those in Model
1. In addition, the oscillation amplitude is steady at high energy,
which is different comparing with Model 1. When neutrino energy is
high enough, parameters $\varepsilon$ and $\zeta$ are much lager
than $\eta$ because $\varepsilon$ and $\zeta$ linearly depend on
neutrino energy. LV parameters $c_{e\mu}^{00}$ and
$c_{\mu\tau}^{00}$ in the numerator and denominator have the same
power in the mixing part, so the values of mixing part are steady at
high energy.

\subsection{Model 3 Both of CPT-Even and CPT-Odd parameters are included in the diagonal term}
\label{sec:model3}

If we assume that $a_{e\mu}^{0}$, $a_{\mu\mu}^{0}$,
$a_{\mu\tau}^{0}$, and $c_{\mu\mu}^{00}$ are nonzero, then
\begin{equation}
\varepsilon=a_{e\mu}^{0}, \qquad \zeta=a_{\mu\tau}^{0}, \qquad
\eta=a_{\mu\mu}^{0}+c_{\mu\mu}^{00}E.
\end{equation}

\begin{figure}[h]
\centering \scalebox{0.7}{\includegraphics[0,0][250,197]{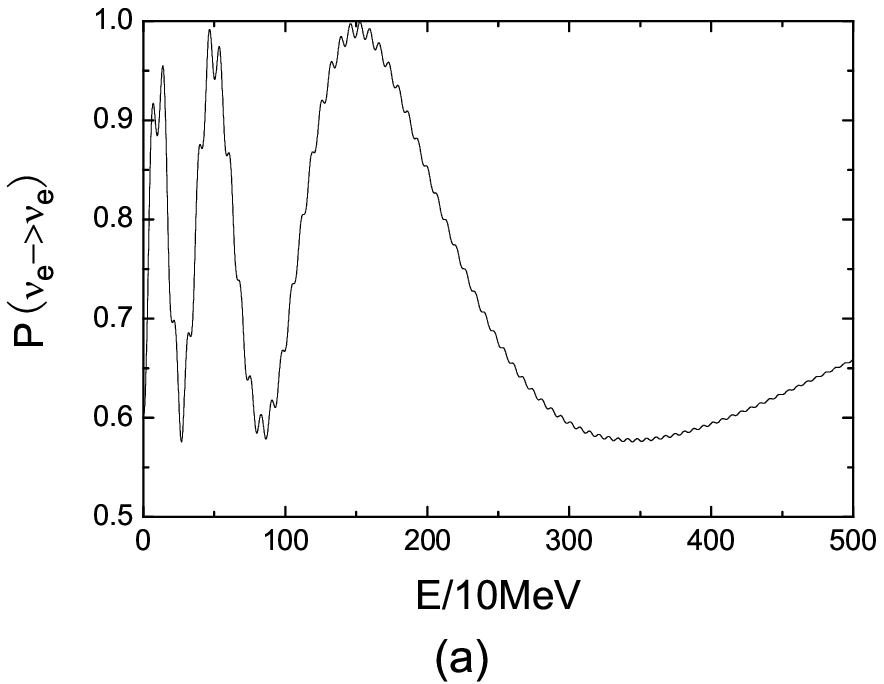}}
\scalebox{0.7}{\includegraphics[0,0][250,197]{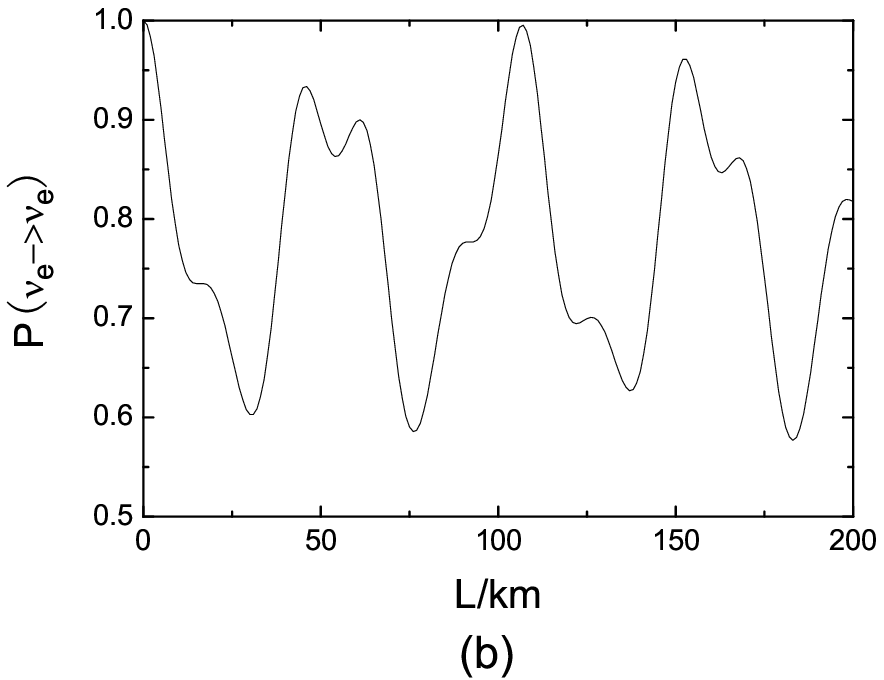}}
\caption{Neutrino oscillation probabilities for $\nu_{e}\to\nu_{e}$.
(a) The left plot shows the oscillation probability as a function of
neutrino energy with a fixed path length $L=180$ km. (b) The right
plot shows the oscillation probability as a function of path length
with a fixed energy $E=10$ MeV.} \label{fig:4}
\end{figure}

\begin{figure}[h]
\centering \scalebox{0.7}{\includegraphics[0,0][250,197]{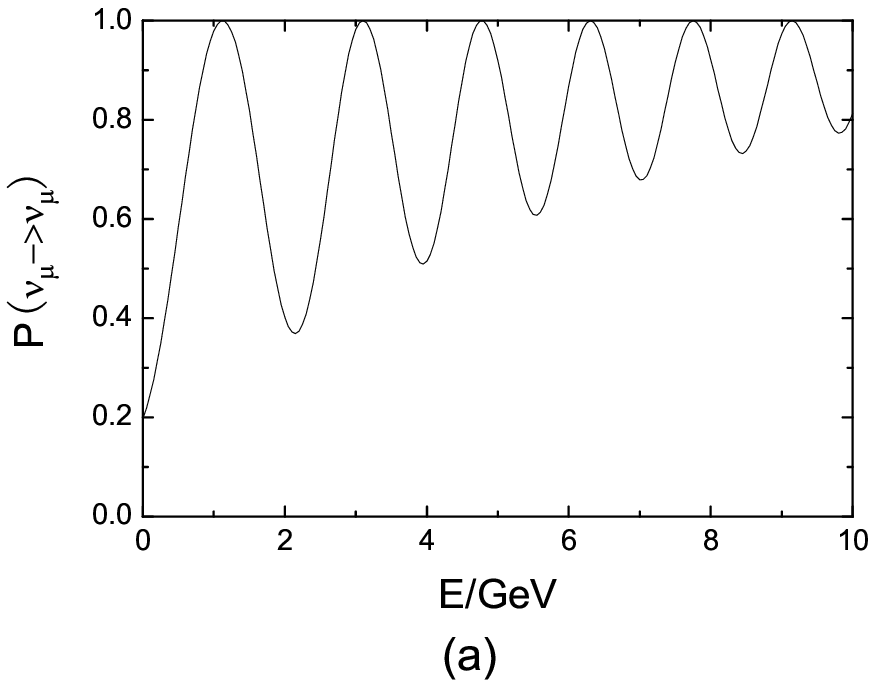}}
\scalebox{0.7}{\includegraphics[0,0][250,197]{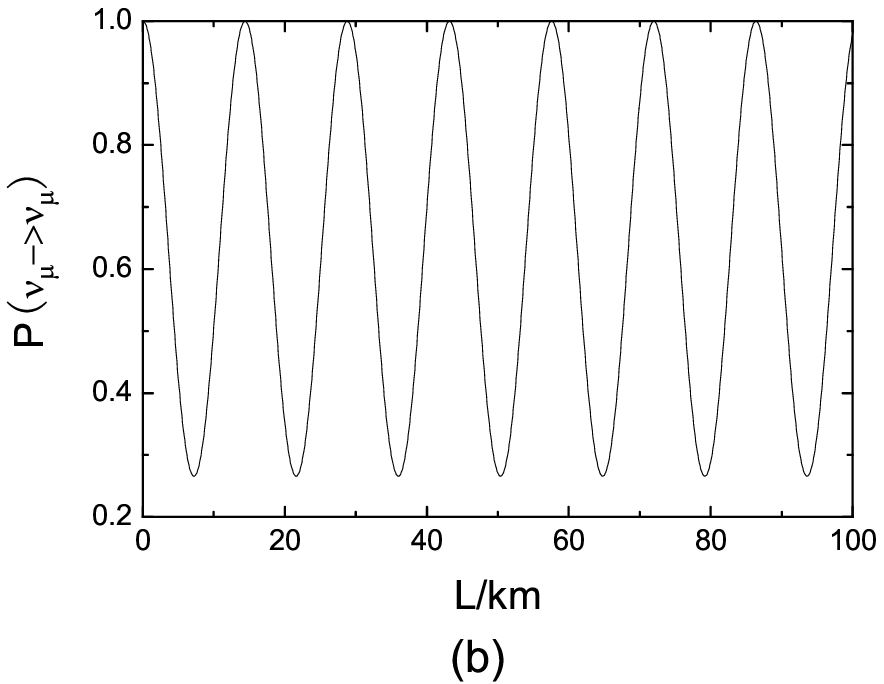}}
\caption{Neutrino oscillation probabilities for
$\nu_{\mu}\to\nu_{\mu}$. (a) The left plot shows the oscillation
probability as a function of neutrino energy with a fixed path
length $L=100$ km. (b) The right plot shows the oscillation
probability as a function of path length with a fixed energy $E=1$
GeV.} \label{fig:5}
\end{figure}

We introduce a new equation $\zeta=x\varepsilon$, then the
constraint of the mixing part $\sin ^{2}2\theta_{\mu e}<0.13$ from
K2K experimental results can be easily satisfied by adjusting the
constant $x$. In our calculation we choose $x=2.7$. Substitute all
these new parameters to Eq.~(\ref{Lv:equ:probability}). Comparing
with KamLAND, K2K and MINOS experimental results, we can figure out
the LV parameters. Figs.~\ref{fig:4} and \ref{fig:5} show the
oscillation probability against neutrino energy $E$ and propagation
length $L$ respectively with LV parameters
$a_{e\nu}^{0}=1.28\times10^{-11}$ eV,
$a_{\mu\mu}^{0}=3.40\times10^{-11}$ eV, and
$c_{\mu\mu}^{00}=1.04\times10^{-20}$.

Figs.~\ref{fig:4} and \ref{fig:5} keep the main characteristics of
Figs.~\ref{fig:1} and \ref{fig:2}. Neutrinos do not have drastic
oscillation at low energy and the oscillation phenomena still exists
at high energy, which is different from the massive neutrino
scenario. Similar to Model 1, the oscillation amplitude varies in
different energy scale. These similarities can be explained by the
fact that the oscillation probability dependence on LV parameters
and neutrino energy have similar form in the two models .

During the above calculation the boost effect has not been
considered. In this work we assumed that the LV parameters are
direction independent in the sun-centered frame.  The experiments
are done on the Earth which is moving around the sun and is also
rotating. The experiments in laboratories therefore see LV
parameters which have direction dependence because there is a boost.
When the long baseline experiments based on the earth are discussed,
the boost effect would cause some influence on the direction
dependent parameters. As the earth moving velocity is much smaller
than the light speed, the boost effect, which is about $10^{-4}$ for
orbit and $10^{-6}$ for rotation respectively, leads to little
impact to the numerical values of the direction independent
parameters ($a^{0}$ and $c^{00}$) in the earth inertial frame. But
the influence on the direction dependent parameters $a^{_{X}}$ and
$c^{_{0X}}$, which are about $10^{-4}$ order comparing to the
corresponding direction independent parameters, needs to be compared
with the upper bounds given in Ref.~\refcite{arXiv:0801.0287}. In
our model 2, the direction dependent parameter $c^{_{0X}}$ induced
by boost effect has the same order comparing with the upper bound in
Ref.~\refcite{arXiv:0801.0287}. Further constraints from future
experiments for direction dependent parameters would be a challenge
for this model. Therefore Ref.~\refcite{arXiv:0801.0287} has put a
strict constraint for the sun-centered directional independent
models built in this work. However, a model with smaller direction
independent parameters, such as our model 1 and model 3, is
consistent with the experimental data.

In addition, LSND data are not used during the calculation of the
three different kinds of models. Substituting the LV parameters
numerical results of these models into the equation
$P(\bar{\nu_{\mu}}\rightarrow\bar{\nu_{e}})$ respectively, we find
that the oscillation probability are much smaller than the
oscillation probability of (0.26$\pm$0.067$\pm$0.045)\% reported by
LSND. So in our work the explanation of the LSND data based on the
two flavor neutrino oscillation is unfavored. While in this work we
assume that only a limited special number of LV parameters are
nonzero, there are still large freedoms to fit experimental results
with Lorentz violation models, e.g., Ref.~\refcite{Katori:2006prd}
explained all class of experimental data, including LSND, with a
massive Lorentz violation model. How to explain LSND data would need
more experimental and theoretical work.

\section{Conclusion}
\label{sec:conclu}

In this paper, we calculated Lorentz violation (LV) contribution to
neutrino oscillation by the effective field theory for LV (Standard
Model Extension). We assume that neutrinos are massless and that
there are only three generations of left-hand neutrinos and
right-handed antineutrinos in nature. Unlike the conventional
massive neutrino scenario, in our calculation neutrino flavor states
are not the mixing states of neutrino mass eigenstates but neutrino
energy eigenstates. In this work, the equation of neutrino
oscillation probabilities given by Eq.~(\ref{Lv:equ:probability}) is
a complete analytic equation and contains hundreds of LV parameters.
In section $3$, we introduced three direction-independent specific
models. Comparing with neutrino experimental results, the upper
bounds on LV parameters can be figured out. We have checked our
calculation results with the experimental results in
Ref.~\refcite{arXiv:0801.0287}. In model 2, the directional
dependent parameters $c_{_{0X}}$ induced by the boost effect are
smaller than the upper bounds given in
Ref.~\refcite{arXiv:0801.0287} but with the same order. And slightly
mediations for the directional independent parameters in model 2 may
lead to its failure. So Ref.~\refcite{arXiv:0801.0287} has put a
strict constraints for the sun-centered directional independent
models built in this work. And a model with  smaller directional
independent parameters, such as model 1 and model 3, is much better
to accord with the experimental data. From
Figs.~\ref{fig:1}$-$\ref{fig:5} we see that neutrinoes do not have
drastic oscillations at low energy and oscillations still exist at
high energy. Furthermore, the oscillation amplitude varies with the
neutrino energy. All these characteristics are different from the
massive neutrino scenario. In addition, the three models in this
paper have different characteristics, too. The main difference lays
in the relationship between oscillation probabilities and neutrino
energy. In models 1 and 3, the oscillation amplitude varies in
different energy scale and goes to zero when the neutrino energy is
high enough. But in model 2, the oscillation amplitude is steady at
high energy. These different models can be tested and constrained in
high energy neutrino experiments.

We also aware the negative report for Lorentz violation from
existing atmospheric neutrino data.
Refs.\cite{Lipari:1999prd,Fogli:1999prd,Maltoni:2004prd} presented
detailed analysis of the zenith angle distribution of atmospheric
neutrino events from Super-Kamiokande underground experiment. The
analysis of super-Kamiokande data disfavors Lorentz-invariance as
the leading source of atmospheric neutrino oscillation. There are
several potential reasons to explain this difficulty. First, Lorentz
violation is not the leading mechanism for atmosphere neutrinos and
the contribution of massive mechanism to neutrino oscillation can
not be neglected at this range. From another point of view, in our
calculation the mixing part is a function of neutrino energy and as
Fig.~\ref{fig:2} shown, the oscillation amplitudes are suppressed at
high energy, which is different from the Lorentz violation model
used in Refs.\cite{Lipari:1999prd,Fogli:1999prd,Maltoni:2004prd}.
(The effect of oscillation suppressing at high neutrino is general
consistent with the conventional massive model, though the
suppressed threshold energy is different between the two kinds of
mechanisms). So these novel effects in our model would produce some
new effect in analyzing atmosphere neutrino from the view of Lorentz
violation.

In the above discussion, we assume that neutrinos are massless and
that Lorentz violation is the only origin of neutrino oscillation.
So we can clearly see what kind of new effect will occur if Lorentz
violation exists in neutrino sector. It is possible that neutrinos
have small mass and both Lorentz violation and the conventional
oscillation mechanism contribute to neutrino oscillation. Then LV
parameter will be further constrained and our numerical calculation
for LV parameters could only be considered as a upper bound for LV.
It is difficult to distinguish between these two kinds of mechanism
at low energy. But high energy experiment is a good avenue to test
Lorentz violation models. In the conventional massive model, the
mixing part is independent of energy and neutrino oscillations
disappear at high energy because the oscillation $P$ is proportional
to $L/E$. But in Lorentz violation models, mixing part is the
function of energy and the oscillation amplitude varies with
neutrino energy. Thus neutrino experiments with energy dependence
may distinguish between the conventional massive neutrino scenario
and the Lorentz violation scenario for neutrino oscillations.

\section*{Acknowledgments}

We acknowledge Zhi-Qiang Guo, Shi-Wen Li, and Zhi Xiao for useful
discussions. This work is partially supported by National Natural
Science Foundation of China (Nos.~10721063, 10575003, 10528510), by
the Key Grant Project of Chinese Ministry of Education (No.~305001),
by the Research Fund for the Doctoral Program of Higher Education
(China).

%\begin{thebibliography}{000} %for 3 digits
%\begin{thebibliography}{00}  %for 2 digits


\begin{thebibliography}{0}    %for 1 digit

\bibitem{Kostelecky:1989prd}
V.~A.~Kosteleck\'y and S.~Samuel, {\it Phys. Rev. D\/} {\bf 39}, 683
(1989).

\bibitem{Ellis:1999gr}
J.~R.~Ellis, N.~E.~Mavromatos and D.~V.~Nanopoulos, gr-qc/9909085;

O.~Bertolami, R.~Lehnert, R.~Potting and A.~Ribeiro, {\it Phys. Rev.
D\/} { \bf 69}, 083513 (2004).

\bibitem{Burgess:2002jhep}
C.~P.~Burgess, J.~Cline, E.~Filotas, J.~Matias and G.~D.~Moore,
 {\it J. High Energy Phys.\/} {\bf 03}, 043 (2002).

\bibitem{Gambini:1999prd}
R.~Gambini and J.~Pullin, {\it Phys. Rev. D\/} {\bf 59}, 124021
(1999).

\bibitem{Douglas:2001rev}
M.~R.~Douglas and N.~A.~Nekrasov, {\it Rev. Mod. Phys\/} {\bf 73},
977 (2001).

\bibitem{Chadha:1983nucl}
S.~Chadha and H.~B.~Nielsen, {\it Nucl. phys. B\/} {\bf 217}, 125
(1983).

\bibitem{Kraus:2002prd}
P.~Kraus and E.~T.~Tomboulis, {\it Phys. Rev. D\/} {\bf 66}, 045015
(2002).

\bibitem{Jenkins:2004prd}
A.~Jenkins, {\it Phys. Rev. D\/} {\bf 69}, 105007 (2004).

\bibitem{Damour:1994Nucl}
T.~Damour and A.~M.~Polyakov, {\it Nucl. phys. B\/} {\bf 423}, 532
(1994).

\bibitem{Arkani:2004jhep}
N.~Arkani-Hamed, H.~C.~Cheng, M.~A.~Luty and S.~Mukohyama, {\it J.
High Energy Phys.\/} {\bf 05}, 074 (2004).

\bibitem{Kostelecky:2003prd}
V.~A.~Kosteleck\'y, R.~Lehnert and M.~J.~Perry, {\it Phys. Rev. D\/}
{\bf 68}, 123511 (2003).

\bibitem{bertolami:2004prd}
O.~Bertolami, R.~Lehnert, R.~Potting and A.~Ribeiro, {\it Phys. Rev.
D\/} {\bf 69}, 083513 (2004) .

\bibitem{Moffat:1993int}
J.~W.~Moffat, {\it Int. J. Mod. Phys. D\/} {\bf 2}, 351 (1993).

\bibitem{Magueijo:2003rept}
J.~Magueijo, {\it Rept. Prog. Phys\/} {\bf 66}, 2025 (2003).

\bibitem{Coleman:1998ph}
S.~R.~Coleman and S.L.~Glashow, hep-ph/9808446.

\bibitem{Coleman:1999prd}
S.R.~Coleman and S.L.~Glashow, {\it Phys. Rev. D\/} {\bf 59}, 116008
(1999).

\bibitem{Kifune:1999astro}
T.~Kifune, {\it Astrophys. J. Lett.\/} {\bf L21}, 518 (1999).

\bibitem{Protheroe:2000plb}
R.J.~Protheroe and H.~Meyer, {\it Phys. Lett. B\/} {\bf 493}, 1
(2000).

\bibitem{Piran:2001ph}
G.~Amelino-Camelia and T.~Piran, {\it Phys. Lett. B\/} {\bf 497},
265 (2001).

\bibitem{Wolfgang:2008ar}
Wolfgang Bietenholz, arxiv:0806.3713v2.

\bibitem{Aharonian:1999as}
HEGRA Collab. (F.~Aharonian {\it et al}.), {\it Astron.
Astrophys.\/} {\bf 349}, 29 (1999).

\bibitem{Greisen:1966prl}
K.~Greisen, {\it  Phys. Rev. Lett.\/} {\bf 16}, 748 (1966).

\bibitem{Zatsepin:1966jetp}
G.~T.~Zatsepin and V.~A.~Kuzmin, {\it  JETP Lett.\/} {\bf 4}, 78
(1966).

\bibitem{Takeda:1998prl}
M.~Takeda {\it et al.}, {\it  Phys. Rev. Lett.\/} {\bf 81}, 1163
(1998).

\bibitem{Xiao:2008ar}
Z.~Xiao, B.-Q.~Ma, {\it  Int. J. Mod. Phys. A\/} {\bf 24}, 1359
(2009).

\bibitem{Abbasi:2007ar}
High Resolution Fly¡¯s Eye Collab. (R.~Abbasi {\it et al}.), {\it
Phys. Rev. Lett.\/} {\bf 100}, 101101 (2008).

\bibitem{Pierre:2007sc}
Pierre Auger Collab. (J. Abraham {\it et al}.), {\it  Science\/}
{\bf 318}, 938 (2007);

The Pierre Auger Collab. (J. Abraham {\it et al}.), {\it  Phys. Rev.
Lett.\/} {\bf 101}, 061101 (2008).

\bibitem{Mattingly:2005living}
D. Mattingly, {\it  Living Rev. Rel.\/}{\bf 8}, 5 (2005).

\bibitem{Mewes:2004hep}
M.~Mewes, hep-ph/0409344.

\bibitem{MNS:1962prog}
Z.~Maki, M.~Nakagawa and S.~Sakata, {\it  Prog. Theor. Phys.\/} {\bf
28}, 870 (1962).

\bibitem{Pontecorvo:1967zh}
B.~Pontecorvo, {\it  Zh. Eksp. Teor. Fiz.\/} {\bf 33}, 549 (1957).

\bibitem{Coleman:1997plb}
S.~R.~Coleman and S.~L.~Glashow, {\it  Phys. Lett. B\/} {\bf 405},
249 (1997).

\bibitem{Barger:2000prl}
V.~Barger, S.~Pakvasa, T.~J.~Weiler and K.~Whisnant, {\it  Phys.
Rev. Lett.\/} {\bf 85}, 5055 (2000).

\bibitem{Kostelecky:2004prd}
V.~A.~Kosteleck\'y and M.~Mewes, {\it  Phys. Rev. D\/} {\bf 69},
016005 (2004).

\bibitem{Kostelecky:2004prd2}
V.~A.~Kosteleck\'y and M.~Mewes, {\it  Phys. Rev. D\/} {\bf 70},
031902 (2004).

\bibitem{Katori:2006prd}
T.~Katori, V.~A.~Kosteleck\'y and R.~Tayloe, {\it  Phys. Rev. D\/}
{\bf 74}, 105009 (2006).

\bibitem{Grassman:2005prd}
Y.~Grossman, C. Kilic, J.~Thaler and G.~E.~Walker, {\it  Phys. Rev.
D\/} {\bf 72}, 125001 (2005).

\bibitem{Arias:2007plb}
P.~Arias, J.~Gamboa, F.~M\'endez, A.~Das and J.~Lopez-Sari\'on, {\it
Phys. Lett. B\/} {\bf 650}, 401 (2007).

\bibitem{Morgan:2007as}
D.~Morgan, E.~Winstanley, J.~Bruuner and L.~F.~Thompson,
arXiv:0705.1897.

\bibitem{Kostelecky:1997prd}
D.~Colladay, V.~A.~Kosteleck\'y, {\it  Phys. Rev. D\/} {\bf 55},
6760 (1997).

\bibitem{Kostelecky:1998prd}
D.~Colladay, V.A.~Kosteleck\'y, {\it  Phys. Rev. D\/} {\bf 58},
116002 (1998).

\bibitem{Kostelecky:2001prd}
V.~A.~Kosteleck\'y, R.~Lehnert, {\it  Phys. Rev. D\/} {\bf 63 },
065008 (2001).

\bibitem{arXiv:0706.1085}
V.~Barger, D.~Marfatia, K.~Whisnant {\it  Phys.Lett.B\/} {\bf 653 },
267 (2007).

\bibitem{MINOS:arxiv}
MINOS Collab. ( P.~Adamson {\it et al}.), {\it Phys. Rev. Lett.\/}
{\bf 101}, 151601 (2008).


\bibitem{Kamland:2003prl}
KamLAND Collab. (K.~Eguchi {\it et al}.), {\it Phys. Rev. Lett.\/}
{\bf 90}, 021802 (2003);

KamLAND Collab. (K.~Eguchi {\it et al}.), {\it Phys. Rev. Lett.\/}
{\bf 94}, 081801 (2003).

\bibitem{MINOS:2006prl}
MINOS Collab. (D.~G.~Michael {\it et al}.), {\it Phys. Rev. Lett.\/}
{\bf 97}, 191801 (2006).

\bibitem{K2K:2005prl}
K2K Collab. (E.~Aliu {\it et al}.), {\it Phys. Rev. Lett.\/} {\bf
94}, 081802 (2005).

\bibitem{LSND:2001prd}
LSND Collab. (A.~Aguilar {\it et al}.), {\it Phys. Rev. D\/} {\bf
64}, 112007 (2001);

LSND Collab. (C.~Athanassopoulos {\it et al}.), {\it Phys. Rev.
Lett.\/} {\bf 81}, 1774 (1998).

\bibitem{MiniBooNE:2007prl}
MiniBooNE Collab. (A.~A.~Aguilar-Arevalo {\it et al}.), {\it Phys.
Rev. Lett.\/} {\bf 98}, 231801 (2007).

\bibitem{K2K:2006prl}
K2K Collab. (S.~Yamamota {\it et al}.), {\it Phys. Rev. Lett.\/}
{\bf 96}, 181801 (2006).

\bibitem{arXiv:0801.0287}
V.~A.~Kosteleck\'y, N.~Russell, arXiv:0801.0287

\bibitem{Lipari:1999prd}
P.~Lipari and M.~Lusignoli, {\it Phys. Rev. D\/} {\bf 60},  013003
(1999).

\bibitem{Fogli:1999prd}
G.~L.~Fogli, E.~Lisi, A.~Marrone and G.~Scioscia, {\it Phys. Rev.
D\/} {\bf 60}, 053006 (1999).

\bibitem{Maltoni:2004prd}
M.~C.~Gonzalez-Garcia and M.~Maltoni, {\it Phys. Rev. D\/} {\bf 70},
033010 (2004).

\end{thebibliography}
\end{document}